\documentclass[12pt]{article} 
\usepackage[pctex32]{graphics}
\usepackage {graphicx}
\usepackage{amsmath,amsxtra,amssymb,latexsym, amscd}
\usepackage[mathscr]{eucal}

\makeindex
\begin{document}
\parindent=1.05cm 
\setlength{\baselineskip}{14truept}
\setcounter{page}{1}
\makeatletter 
\title{}
\date{} 
\maketitle
\pagestyle{plain}
\pagestyle{myheadings}
\markboth{\footnotesize 
Pham Thuc Tuyen, Do Quoc Tuan  
}{\footnotesize  
On the string solution in the SUSY - Skyrme model
}
\begin{center}
\vskip-3cm 
{\large \bf 
On the string solution in the SUSY - Skyrme model
}
\end{center}
\vskip0.1cm 
\centerline{
Pham Thuc Tuyen \footnote {Email: tuyenpt@coltech.vnu.vn}
}
\vskip0.1cm 
\begin{center}{\it 
Department of Theoretical Physics, Hanoi University of Science, 334 Nguyen Trai, Thanh Xuan, Ha Noi, Viet Nam.
}
\end{center}
\vskip0.1cm 
\centerline{
Do Quoc Tuan \footnote {Email: do.tocxoan@gmail.com. Associate address after February 2008: Department of Computing Physics, Hanoi University of Science, 334 Nguyen Trai, Thanh Xuan, Ha Noi} 
}
\vskip0.1cm 
\begin{center}{\it
 
Department of Theoretical Physics, Hanoi University of Science, 334 Nguyen Trai, Thanh Xuan, Ha Noi, Viet Nam.
}
\end{center}
\vskip0.4cm 
\leftskip1.05cm 

{\small \noindent {\bf Abstract}: In this paper, we have found the string solution in the SUSY Skyrme model. Moreover, the mechanics of decay of SUSY - string was discussed. }
\vskip0.4cm 
\leftskip1.05cm 
{\it Keywords} : String, SUSY, Skyrme model. 
\leftskip0cm 
\vskip0.4cm \vskip0.4cm 
\section{Introduction}

String - like solution firstly obtained \cite{J} from equaton of motion of the Skyrme model with a pion's mass term by A. Jackson. This string-like solution may be closely related to QCD string \cite{J}. This string solution is unstable and it may decay by emitting pions \cite{J}. During the decay a baryon current flows along the string, producing half a baryon and half an antibaryon. Long strings can decay via many different decay models, some producing baryon-antibaryon pairs \cite{J}. Recently, M. Nitta and M. Skiiki constructed non-topological string solutions with $U(1)$ Noether charge in the Skyrme model with a pion mass term. And they also showed that this string were not stabilized by $U(1)$ rotation and decay into baryon-antibaryon pairs or mesons in the same way as the string without the charge. They showed that a rotating confugutation would reduce its rotational energy by emmiting pions. When this string becomes longer than $\pi/{\hat m_\pi}$, it will decay by emitting pions \cite{NS}.
In this paper we want to connect idea of the string \cite{J} to the supersymmetric skyrme model proposed by E. A. Bershoeff at el. \cite{BN}. This way may give us the SUSY string solution (superstring) and a mechanics of decay squark-antisquark, baryon-antibaryons, etc.
\leftskip0cm 
\vskip0.4cm \vskip0.4cm 
\section{The SUSY string solution}
\vskip0.2cm 
Let us consider the lagrangian with a mass of pion \cite{AN}
\begin{equation}
{\cal L} =  - \frac{{F_\pi ^2 }}{{16}}Tr\left( {\partial _\mu  U^\dag  \partial ^\mu  U} \right) + \frac{1}{{32e^2 }}Tr\left( {\left[ {U^\dag  \partial _\mu  U,U^\dag  \partial _\nu  U} \right]^2 } \right) + \frac{1}{8}m_\pi ^2 F_\pi ^2 Tr\left[ {1 - U} \right].
\end{equation}
In terms of complex scalars $A_i$ \cite{BN}, this equation can be rewritten as
\begin{equation}
{\cal L} = {\cal L}_{susy}  + \frac{1}{8}m_\pi ^2 F_\pi ^2 \left[ {\bar A_1  + A_1  - 2} \right].
\end{equation}
Now, let us consider the rotating soliton developed by M. Nitta at el. \cite{NS} from the original soliton constructed by A. Jackson \cite{J}
\begin{equation}
U = \left[ {\begin{array}{*{20}c}
   {\cos f\left( r \right)} & {i\sin f\left( r \right)e^{ - i\left( {\theta  + \alpha \left( t \right)} \right)} }  \\
   {i\sin f\left( r \right)e^{i\left( {\theta  + \alpha \left( t \right)} \right)} } & {\cos f\left( r \right)}  \\
\end{array}} \right]
\end{equation}
\begin{equation}
\to A_1  = \cos f\left( r \right);A_2  = i\sin f\left( r \right)e^{i\left( {\theta  + \alpha \left( t \right)} \right)} ,
\end{equation}
in the cylindrical coordinate system with the metric
\begin{equation}
ds^2  =  - dt^2  + dz^2  + dr^2  + r^2 d\theta ^2 .
\end{equation}
Substituting this solution into Lagrangian (2), we obtain the string tension (see more in the appendix)
\begin{center}
${\cal E} = \int {4\pi r^2 } dr\left\{ {\frac{{ - f_\pi ^2 }}{8}\left[ {\left( { - \dot \alpha ^2  + \frac{1}{{r^2 }}} \right)\sin ^2 f + f'^2 } \right] + } \right.$
\end{center}
\begin{center}
$ + \frac{1}{{8e^2 }}\left[ {\left( {a - b} \right)\left( {\left( {\dot \alpha ^4  - 2\frac{{\dot \alpha ^2 }}{{r^2 }} + \frac{1}{{r^4 }}} \right)\sin ^4 f + 2f'^2 \sin ^2 f\left( {\dot \alpha ^2  + \frac{1}{{r^2 }}} \right) + f'^4 } \right) + } \right.$
\end{center}
\begin{equation}
\left. {\left. { + b\left( {\left( {\dot \alpha ^4  + \ddot \alpha ^2  + 1} \right)\sin ^2 f + f''^2  + f'^4 } \right)} \right] + \frac{1}{4}m_\pi ^2 f_\pi ^2 \left( {1 - \cos f} \right)} \right\} .
\end{equation}
Setting the dimensionless variable $\rho  = f_\pi  er \equiv \gamma r$, and taking the variations of $f\left( \rho  \right)$ in the string tension $\delta _f {\cal E} = 0$, we have the Euler-Lagrange equation
\begin{center}
$\frac{\partial }{{\partial \rho }}\frac{{\delta {\cal E}}}{{\delta f'}} - \frac{{\delta {\cal E}}}{{\delta f}} = 0$
\end{center}
\begin{center}
$ \to f''\left[ { - 2\rho ^2  + 4\rho ^2 (a-b)\sin ^2 f\left( {\frac{{\dot \alpha ^2 }}{{\gamma ^2 }} + \frac{1}{{\rho ^2 }}} \right) + 12\rho ^2 af'^2 } \right] + $
\end{center}
\begin{center}
$ + f'^3 \left[ {8\rho a - 2\sin 2f\left( {\frac{{\dot \alpha ^2 }}{{\gamma ^2 }} + \frac{1}{{\rho ^2 }}} \right)} \right] + 4f'^2 \rho ^2 (a-b)\sin 2f\left( {\frac{{\dot \alpha ^2 }}{{\gamma ^2 }} + \frac{1}{{\rho ^2 }}} \right) + $
\end{center}
\begin{center}
 $+ f'\left[ { - 4\rho  + 8\rho \left( {a - b} \right)\sin ^2 f\left( {\frac{{\dot \alpha ^2 }}{{\gamma ^2 }} + \frac{1}{{\rho ^2 }}} \right) - 2\sqrt \rho  \left( {a - b} \right)\sin ^2 f - } \right.$
\end{center}
\begin{center}
$\left. {-2\rho ^2 \left( {a - b} \right)\left( {\frac{{\dot \alpha ^4 }}{{\gamma ^4 }} - \frac{{2\dot \alpha ^2 }}{{\gamma ^2 \rho ^2 }} + \frac{1}{{\rho ^4 }}} \right)\sin 2f\sin ^2 f} \right] + $
\end{center}
\begin{equation}
+ \rho ^2 \left[ \frac{-b}{{\gamma ^4 }}\left( {\ddot \alpha ^2  + \dot \alpha ^4  + 1} \right)+{\frac{{\dot \alpha ^2 }}{{\gamma ^2 }}+\frac{1}{{\rho ^2 }}}\right]\sin 2f - 2\frac{{m_\pi ^2 }}{{\gamma ^2 }}\rho ^2 \sin f = 0 .
\end{equation}
For the finiteness and regularity conditions of the string tension, one requires
\begin{equation}
f\left( 0 \right) = n\pi ; f\left( \infty  \right) = 0 ,
\end{equation}
where $n$ is a positive integer. This field equation can be solved numerically with two boundary conditions (8). 
\subsection{ the case of $b=0$}
In the configuration of rotating soliton $\alpha \left( t \right)$ is angular rotation of soliton in time \cite{J,NS} thus the quantity $\hat \omega $ defined as $\dot \alpha  = \frac{\omega }{\gamma } \equiv \hat \omega $ will be understood as an angular velocity of rotating soliton. To obtain the asymptotic form of the profile $f\left( \rho  \right)$ as $\rho  \to \infty $ we need linearize the field equation (7). Setting $f = \delta f$, we get
\begin{equation}
\rho ^2 \delta f'' + 2\rho \delta f' + \left[ {1 - \rho ^2m^2 } \right]\delta f -  = 0 ,
\end{equation}
where $m^2  = \left( {\frac{{m_\pi ^2 }}{{\gamma ^2 }} - \hat \omega ^2 } \right)$ .
This is the Bessel equation, to obtain solutions of Bessel function, we require 
\begin{equation}
0 < \hat \omega  < \hat m_\pi,
\end{equation}
where $\hat m_\pi=\frac{{m_\pi  }}{\gamma }$ .
\\
Let us mention in ref. \cite{NS}, the condition for angular velocity was obtained as $0 < \hat \omega  < \frac{{\hat m_\pi  }}{{\sqrt 2 }}$ and they shown the mechanics of emitting pions of string when $\hat \omega $ increases over the critical value $\hat \omega _ +   = \hat m_\pi  /\sqrt 2$. However, in SUSY case, our critical value is $\hat \omega _ + ^{susy}  = \hat m_\pi  = \sqrt 2 \hat \omega _ +  $. In cases of critical values are larger than this critical value, the SUSY - String solution can be emitted into baryons - sbaryons pairs and pions - spions. 
\subsection{ the case of $b \ne 0$}
For this case, we have the equation
\begin{equation}
\delta f'' + \frac{2}{\rho }\delta f' + \frac{1}{{\rho ^2 }}\delta f + \left[ { - \frac{b}{{\gamma ^2 }}\left( {\frac{{d\hat \omega }}{{dt}}} \right)^2  - b\hat \omega ^4  + \hat \omega ^2  - \frac{b}{{\gamma ^4 }} - \hat m_\pi ^2 } \right]\delta f = 0 .
\end{equation}
We now consider an angular velocity of soliton is constant, means $\frac{{d\omega }}{{dt}} = 0$. Similarly to a case of $b=0$, we require 
\begin{equation}
\left[ - b\hat \omega ^4  + \hat \omega ^2  - c \right] > 0 ,
\end{equation}
where $c = \frac{b}{{\gamma ^4 }} +\hat m_\pi ^2  $.
Following conditions for $b$ and $\omega$ are given as
\begin{equation}
- \gamma ^2 m_\pi ^2  < b < 0 \to 0 < \omega  < \sqrt {x_2 },
\end{equation} 
\begin{equation}
0 < b < \frac{{\gamma ^2 \left( { - 1 + \sqrt {1 + m_\pi ^2 } } \right)}}
{2}\to \sqrt {x_1 }  < \omega  < \sqrt {x_2 }, 
\end{equation}
where $x_1  = \frac{{ - 1 - \sqrt {1 - 4bc} }}{{ - 2b}},x_2  = \frac{{ - 1 + \sqrt {1 - 4bc} }}{{ - 2b}}$.
Eqs (13), (14) show us areas of $\omega$ in which SUSY string may be decay. This is a different point between non-SUSY string \cite{J, NS} and SUSY string.   

\leftskip0cm 
\vskip0.4cm \vskip0.4cm 
\section {Appendix}
\subsection {Recall the SUSY Skyrme model}
Let us consider the SUSY Lagrangian Skyrme \cite{BN, SSO}
\begin{equation}
{\cal L} =  - \frac{{f_\pi ^2 }}{{16}}Tr\left( {\partial _\mu  U^\dag  \partial ^\mu  U} \right) + \frac{1}{{32e^2 }}Tr\left( {\left[ {U^\dag  \partial _\mu  U,U^\dag  \partial _\nu  U} \right]^2 } \right) ,
\end{equation}
where $U$ is can $SU(2)$ matrix, $f_\pi$ is the pion decay constant, $e$ is a free parameter.
The ordinary derivatives in the Lagrangian density (15) is replaced by the covariant derivatives
\begin{equation}
\partial _\mu  U \to D_\mu  U = \partial _\mu  U - iV_\mu  U\tau _3 ,
\end{equation}
Eq (15) becomes as
\begin{equation}
{\cal L} =  - \frac{{f_\pi ^2 }}{{16}}Tr\left( {D_\mu  U^\dag  D^\mu  U} \right) + \frac{1}{{32e^2 }}Tr\left( {\left[ {U^\dag  D_\mu  U,U^\dag  D_\nu  U} \right]^2 } \right) .
\end{equation}
Eq (17) is invariant under the local $U\left( 1 \right)_R $ and the global $SU\left( 2 \right)_L $ transformations
\begin{center}
$U\left( r \right) \to AU\left( r \right)e^{i\lambda \left( r \right)\tau _3 } ,A \in SU\left( 2 \right)_L $ ,
\end{center}
\begin{equation}
V_\mu  \left( r \right) \to V_\mu  \left( r \right) + \partial _\mu  \lambda \left( r \right) .
\end{equation}
where the gauge field $V_\mu  \left( r \right)$  is defined as
\begin{equation}
V_\mu   =  - \frac{i}{2}Tr\left( {U^\dag  \partial _\mu  U\tau _3 } \right) .
\end{equation}
One parametrizes the $SU(2)$ matrix $U$ in terms of the complex scalars $A_i$
\begin{equation}
U\left( r \right) = \left( {\begin{array}{*{20}c}
   {A_1 } & { - A_2^* }  \\
   {A_2 } & {A_1^* }  \\
\end{array}} \right),
\end{equation}
where $\bar A^i A_i  = A_1^* A_1  + A_2^* A_2  = 1$ .
Eq (16) can be rewritten as
\begin{center}
$D_\mu  A_i  = \left( {\partial _\mu   - iV_\mu  } \right)A_i $ ,
\end{center}
\begin{equation}
D_\mu  \bar A_i  = \left( {\partial _\mu   + iV_\mu  } \right)\bar A_i 
\end{equation}
and the new form of gauge field is
\begin{equation}
V_\mu  \left( r \right) =  - \frac{i}{2}\bar A^i \mathord{\buildrel{\lower3pt\hbox{$\scriptscriptstyle\leftrightarrow$}} 
\over \partial } A_i . 
\end{equation}
Finally, we obtain Lagrangian in terms of complex scalars $A_i$
\begin {equation}
{\cal L} =  - \frac{{f_\pi ^2 }}{8}\bar D_\mu  \bar AD^\mu  A - \frac{1}{{16e^2 }}F_{\mu \nu }^2 , 
\end{equation}
where $F_{\mu \nu } \left( V \right) = \partial _\mu  V_\nu   - \partial _\nu  V_\mu  $
One supersymmetrised Skyrme model by extending  $A_i$  to chiral scalar multiplets $\left( {A_i ,\psi _{\alpha i} ,F_i } \right)$ $\left( {i,\alpha  = 1,2} \right)$ and the vector $V_\mu  \left( x \right)$ to real vector multiplets $\left( {V_\mu  ,\lambda _\alpha  ,D} \right)$. Here, the fields $F_i$ are complex scalars, $D$ is real scalar, $\psi _{\alpha i} $, $\lambda _\alpha  $ are Majorana two-component spinors. $\psi _{\alpha i} $ corresponds to a left-handed chiral spinor, $\bar \psi ^{\alpha i}  = \left( {\psi _i^\alpha  } \right)^* $ corresponds to a right-handed one. The SUSY Lagrangian density is given by
\begin{center}
${\cal L}_{susy}  = \frac{{f_\pi ^2 }}{8}\left[ { - D^\mu  \bar A^i D_\mu  A_i  - \frac{1}{2}i\bar \psi ^{\dot \alpha i} \left( {\sigma _\mu  } \right)_{\alpha \dot \alpha } \mathord{\buildrel{\lower3pt\hbox{$\scriptscriptstyle\leftrightarrow$}} 
\over D} ^\mu  \psi _i^\alpha   + \bar F^i F_i  - } \right .
$
\end{center}
\begin{center}
$\left. { - i\bar A^i \lambda ^\alpha  \psi _{\alpha i}  + iA_i \bar \lambda ^{\dot \alpha } \bar \psi _{\dot \alpha }^i  + D\left( {\bar A^i A_i  - 1} \right)} \right] + $
\end{center}
\begin{equation}
+ \frac{1}{{8e^2 }}\left[ { - \frac{1}{2}F_{\mu \nu }^2  - i\bar \lambda ^{\dot \alpha } \left( {\sigma ^\mu  } \right)_{\dot \alpha }^\alpha  \partial _\mu  \lambda _\alpha   + D^2 } \right] .
\end{equation}
This Lagrangian is invariant under the following set of supersymmetric transformations
\begin{equation}
\delta A_i  =  - \varepsilon ^\alpha  \psi _{\alpha i} ,
\end{equation}
\begin{equation}
\delta \psi _{\alpha i}  =  - i\bar \varepsilon ^{\dot \alpha } \left( {\sigma ^\mu  } \right)_{\alpha \dot \alpha } D_\mu  A_i  + \varepsilon _\alpha  F_i ,
\end{equation}
\begin{equation}
\delta F_i  =  - i\bar \varepsilon ^{\dot \alpha } \left( {\sigma ^\mu  } \right)_{\dot \alpha }^\alpha  D_\mu  \psi _{\alpha i}  - i\bar \varepsilon ^{\dot \alpha } A_i \bar \lambda _{\dot \alpha } ,
\end{equation}
\begin{equation}
\delta V_\mu   =  - \frac{1}{2}i\left( {\sigma _\mu  } \right)^{\alpha \dot \alpha } \left( {\bar \varepsilon _{\dot \alpha } \lambda _\alpha   + \varepsilon _\alpha  \bar \lambda _{\dot \alpha } } \right) ,
\end{equation}
\begin{equation}
\delta \lambda _\alpha   = \varepsilon ^\beta  \left( {\sigma ^{\mu \nu } } \right)_{\beta \alpha } F_{\mu \nu }  + i\varepsilon _\alpha  D ,
\end{equation}
\begin{equation}
\delta D = \frac{1}{2}\left( {\sigma ^\mu  } \right)_{\alpha \dot \alpha } \partial _\mu  \left( {\bar \varepsilon ^{\dot \alpha } \lambda ^\alpha   - \varepsilon ^\alpha  \bar \lambda ^{\dot \alpha } } \right) .
\end{equation}
The field equation and their supersymmetric transformations lead to the following constraints
\begin{equation}
\bar A^i A_i  = 0 ,
\end{equation}
\begin{equation}
\bar A^i \psi _{\alpha i}  = 0 ,
\end{equation}
\begin{equation}
\bar A^i F_i  = 0 ,
\end{equation}
and following algebraic expressions for
\begin{equation}
V_\mu   =  - \frac{1}{2}\left\{ {i\bar A^i \mathord{\buildrel{\lower3pt\hbox{$\scriptscriptstyle\leftrightarrow$}} 
\over \partial } _\mu  A_i  + \left( {\sigma _\mu  } \right)^{\alpha \dot \alpha } \bar \psi _{\dot \alpha }^i \psi _{\alpha i} } \right\} ,
\end{equation}
\begin{equation}
\lambda _\alpha   =  - i\left\{ {\bar F^i \psi _{\alpha i}  + i\left( {\sigma ^\mu  } \right)_{\alpha \dot \alpha } \left( {D_\mu  A_i } \right)\bar \psi ^{\dot \alpha i} } \right\} ,
\end{equation}
\begin{equation}
D = D^\mu  \bar A^i D_\mu  A_i  + \frac{1}{2}i\bar \psi ^{\dot \alpha i} \left( {\sigma ^\mu  } \right)_{\alpha \dot \alpha } \left( {\mathord{\buildrel{\lower3pt\hbox{$\scriptscriptstyle\leftrightarrow$}} 
\over D} _\mu  \psi _i^\alpha  } \right) - \bar F^i F_i .
\end{equation}
To obtain the minimum of SUSY extension, one set $\psi _{\alpha i}  = F_i  = 0$. Therefore, Eq (24) becomes as
\begin{equation}
{\cal L}_{susy}  =  - \frac{{f_\pi ^2 }}{8}\bar D^\mu  \bar AD_\mu  A + \frac{1}{{8e^2 }}\left[ { - \frac{1}{2}F_{\mu \nu }^2  + \left( {\bar D^\mu  \bar AD_\mu  A} \right)^2 } \right] .
\end{equation}
However, there is another four-derivatives term of $A_i$, one gave the general form of SUSY Lagrangian 
\begin{equation}
\mathcal{L}_{susy}  =  - \frac{{f_\pi ^2 }}
{8}\bar D^\mu  \bar AD_\mu  A + \frac{1}
{{8e^2 }}\left\{ {a\left[ { - \frac{1}
{2}F_{\mu \nu }^2  + \left( {\bar D^\mu  \bar AD_\mu  A} \right)^2 } \right] + b\left\{ {\square \bar A\square A - \left( {\bar D^\mu  \bar AD_\mu  A} \right)^2 } \right\}} \right\} ,
\end{equation}
where $a$, $b$ are constants,  $\square  = D^\mu  D_\mu  $ is the gauge covariant $D'alembertian$. 

\subsection {Some main results for Eq (6)}

\vskip0.2cm
{\bf *	The first term}

We have
\begin{equation}
D_\mu  A_i  = \left( {\partial _\mu   - iV_\mu  } \right)A_i  = \left[ {\partial _\mu   + \frac{1}
{2}\left( {\partial _\mu  \bar A^i } \right)A_i  + \frac{1}
{2}\bar A^i \left( {\partial _\mu  A_i } \right)} \right]A_i ,
\end{equation}
\begin {equation}
\bar D^\mu  \bar A^i  = \left( {\partial ^\mu   + iV^\mu  } \right)\bar A^i  = \left[ {\partial ^\mu   - \frac{1}
{2}\left( {\partial ^\mu  \bar A^i } \right)A_i  - \frac{1}
{2}\bar A^i \left( {\partial ^\mu  A_i } \right)} \right]\bar A^i .
\end{equation}
Inserting forms of $A_i$  (4), let us final results
\begin{center}
$D_0 A_i  =  - \dot \alpha \sin fe^{i\left( {\theta  + \alpha } \right)} $ ; $D_1 A_i  = 0$ ; $D_2 A_i  = i\cos ff'e^{i\left( {\theta  + \alpha } \right)} $; $D_3 A_i  =  - \sin fe^{i\left( {\theta  + \alpha } \right)} $ ,
\end{center}
\begin{center}
$\bar D^0 \bar A_i  = \dot \alpha \sin fe^{ - i\left( {\theta  + \alpha } \right)} $; $\bar D^1 \bar A_i  = 0$; $\bar D^2 \bar A_i  =  - i\cos ff'e^{ - i\left( {\theta  + \alpha } \right)} $; $\bar D^3 \bar A_i  =  - \sin fe^{ - i\left( {\theta  + \alpha } \right)} $ .
\end{center}
\begin{equation}
 \Rightarrow \bar D^\mu  \bar A_i D_\mu  A_i  = \left( { - \dot \alpha ^2  + 1} \right)\sin ^2 f + f'^2 .
\end{equation}

\vskip0.2cm
{\bf *  The second term}

We have
\begin{equation}
\left( {F_{\mu \nu } } \right)^2  = \left( {\partial _1 V_2  - \partial _2 V_1 } \right)\left( {\partial ^1 V^2  - \partial ^2 V^2 } \right) ,
\end{equation}
in terms of forms of $A_i$ (4), we obtain final results
\begin{center}
$F_{12}^2  = F_{21}^2  = 0$; $F_{13}^2  = F_{31}^2  = 0$; $F_{23}^2  = F_{32}^2  = 0$ .
\end{center}
\begin{equation}
 \Rightarrow \left( {F_{\mu \nu } } \right)^2  = 0
\end{equation}

\vskip0.2cm
{\bf *  The third term}

We have
\begin{center}
$\square  = D^\mu  D_\mu   = \left( {\partial ^\mu   - iV^\mu  } \right)\left( {\partial _\mu   - iV_\mu  } \right)$
\end{center}
\begin{equation}
 = \partial ^\mu  \partial _\mu   - i\partial ^\mu  V_\mu   - iV^\mu  \partial _\mu   - V^\mu  V_\mu = \partial ^\mu  \partial _\mu  
\end{equation}
Or 
\begin{equation}
\square \bar A\square A = \partial _0^2 \bar A^i \partial _0^2 A_i  + \partial _1^2 \bar A^i \partial _1^2 A_i  + \partial _2^2 \bar A^i \partial _2^2 A_i  + \partial _3^2 \bar A^i \partial _3^2 A_i .
\end{equation}
Final results are
\begin{equation}
\partial _0^2 \bar A^i \partial _0^2 A_i  = \cos ^2 f\left( {f'^4 } \right) + 2\sin 2f\left( {f'^2 } \right)f'' + \sin ^2 f\left( {f''^2 } \right) + \sin ^2 f\left( {\ddot \alpha ^2  + \dot \alpha ^4 } \right) ,
\end{equation}
\begin{equation}
\partial _1^2 \bar A^i \partial _1^2 A_i  = 0 ,
\end{equation}
\begin{equation}
\partial _2^2 \bar A^i \partial _2^2 A_i  = \sin ^2 f\left( {f'^4 } \right) - 2\sin 2f\left( {f'^2 } \right)f'' + \cos ^2 f\left( {f''^2 } \right) ,
\end{equation}
\begin{equation}
\partial _3^2 \bar A^i \partial _3^2 A_i  = \sin ^2 f .
\end{equation}
\begin{equation}
\Rightarrow \square \bar A\square A = \sin ^2 f\left[ {\dot \alpha ^4  + \ddot \alpha ^2  + 1} \right] + f''^2  + f'^4 
\end{equation}
\section{Conclusion}
In this paper, we have performed analytic calculations, new results were found. In near future, they will be computed clearly by numerical methods, this way will give us the brilliant picture of mechanics of SUSY string's decay.

\begin{center}
\bf ACKNOWLEDGMENT 
\end{center}
We would like to thank Department of Theoretical Physics because of helps for us. One of us (DQT) want to thank Department of Computing Physics, HUS for giving me good conditions of working.

\end{document}